\documentclass[aps,prl,twocolumn,amsmath,amssymb,superscriptaddress,showpacs,showkeys]{revtex4}
\usepackage{dcolumn}
\usepackage{graphicx,color}
\usepackage{textcomp}

\begin {document}
  \newcommand {\nc} {\newcommand}
  \nc {\beq} {\begin{eqnarray}}
  \nc {\eeq} {\nonumber \end{eqnarray}}
  \nc {\eeqn}[1] {\label {#1} \end{eqnarray}}
  \nc {\eol} {\nonumber \\}
  \nc {\eoln}[1] {\label {#1} \\}
  \nc {\ve} [1] {\mbox{\boldmath $#1$}}
  \nc {\ves} [1] {\mbox{\boldmath ${\scriptstyle #1}$}}
  \nc {\mrm} [1] {\mathrm{#1}}
  \nc {\half} {\mbox{$\frac{1}{2}$}}
  \nc {\thal} {\mbox{$\frac{3}{2}$}}
  \nc {\fial} {\mbox{$\frac{5}{2}$}}
  \nc {\la} {\mbox{$\langle$}}
  \nc {\ra} {\mbox{$\rangle$}}
  \nc {\etal} {\emph{et al.}}
  \nc {\eq} [1] {(\ref{#1})}
  \nc {\Eq} [1] {Eq.~(\ref{#1})}
  \nc {\Ref} [1] {Ref.~\cite{#1}}
  \nc {\Refc} [2] {Refs.~\cite[#1]{#2}}
  \nc {\Sec} [1] {Sec.~\ref{#1}}
  \nc {\chap} [1] {Chapter~\ref{#1}}
  \nc {\anx} [1] {Appendix~\ref{#1}}
  \nc {\tbl} [1] {Table~\ref{#1}}
  \nc {\fig} [1] {Fig.~\ref{#1}}
  \nc {\ex} [1] {$^{#1}$}
  \nc {\Sch} {Schr\"odinger }
  \nc {\flim} [2] {\mathop{\longrightarrow}\limits_{{#1}\rightarrow{#2}}}
  \nc {\textdegr}{$^{\circ}$}
  \nc {\inred} [1]{\textcolor{red}{#1}}
  \nc {\inblue} [1]{\textcolor{blue}{#1}}
\nc {\IR} [1] {\textcolor{red}{#1}}
\nc {\IB} [1] {\textcolor{blue}{#1}}
  \nc {\efbcom}[1]{\textcolor{blue}{\bfseries\itshape #1}}
  \nc {\GK} {\ensuremath{\mathrm{\,GK}}}
  \nc {\Msun} {\ensuremath{\,M_{\odot}}}
\title{Low-temperature triple-alpha rate in a full three-body model}
\author{N.B. Nguyen}
\email{nguyenn@nscl.msu.edu}
\affiliation{National Superconducting Cyclotron Laboratory
and Department of Physics and Astronomy,
Michigan State University, East Lansing, Michigan 48824, USA}
\author{F.~M.~Nunes}
\email{nunes@nscl.msu.edu}
\affiliation{National Superconducting Cyclotron Laboratory
and Department of Physics and Astronomy,
Michigan State University, East Lansing, Michigan 48824, USA}
\author{I.J. Thompson}
\affiliation{Lawrence Livermore National Laboratory, L-414,
 Livermore, California 94551, USA}
\author{E.F. Brown}
\affiliation{National Superconducting Cyclotron Laboratory
and Department of Physics and Astronomy,
Michigan State University, East Lansing, Michigan 48824, USA}
\affiliation{Joint Institute for Nuclear Astrophysics}
\date{\today}
\begin{abstract}
A new three-body method is used to compute the rate of the triple-alpha
capture reaction which is the primary source of $^{12}$C in stars. In this work, we combine the
Faddeev hyperspherical harmonics and the R-matrix method to obtain a full solution
to the three-body $\alpha+\alpha+\alpha$ continuum. Particular attention is paid to the
long range effects caused by the pairwise Coulomb interactions.
The new rate agrees with the NACRE rate for temperatures greater than $0.07$ GK,
but a large enhancement  at lower temperature is found ($\approx 10^{12}$ at $0.02$ GK). Our results are
compared to previous calculations where additional approximations were made. We show that the new rate does not significantly change the evolution of stars around one solar mass. In particular, such stars still undergo a red-giant phase consistent with observations, and no significant differences are found in the final white dwarfs.
\end{abstract}
\maketitle

%%%%%%%%%%%%%%%%%%%%%%%%%%%%%%%%%%%%%%%%%%%%%%%%%%%%%%%%%%%%%%%%%%%%%%%%%%%%%%%%%%%%%% introduction

The triple-alpha reaction is key to the production of elements with mass $A \geq 12 $. It plays a role in
many astrophysical processes, from stellar evolution to explosive scenarios. Since the early days when Fred Hoyle postulated that there
should be a $0^+$ resonance close to the $3\alpha$ threshold to justify the observed abundances
of $^{12}$C in stars \cite{hoyle54}, a state experimentally confirmed by Dunbar {\it et al.} \cite{dunbar53},
the accepted mechanism for the triple-alpha capture has been sequential: $\alpha+\alpha \rightarrow \mathrm{^8 Be}(0^+_1)$
followed by $\mathrm{^8Be}(0^+_1)+\alpha \rightarrow \mathrm{^{12}C}(0^+_2)$.
However, if the energy of the alpha particles in
the stellar environment is insufficient to populate the narrow resonances, a direct three-body capture becomes the favored path.
The most recent compilation of charge-particle induced thermonuclear rates (NACRE) \cite{nacre} extrapolates the sequential
model to low energy, an approach that should be questioned \cite{ogata09,garrido11}. Recent theoretical works
\cite{ogata09,garrido11} predict enhancements in the triple-alpha rate by 7 to 20 orders of magnitude when compared
to NACRE, for temperatures around 0.02 GK. Because the triple-alpha rate, even at these low temperatures,  can
have dramatic and widespread effects in astrophysics, it is critical to resolve the large discrepancies found between
\cite{nacre}, \cite{ogata09} and \cite{garrido11}. This is the aim of the present work.

The Hoyle resonance dominates the triple-alpha rate at $T \sim 0.1$ GK, while other  resonances play a role at $T\sim 1$ GK
relevant in explosive environments. At these high temperatures, a $^{12}$C($0^+_3$) state around $3$ MeV above the $3\alpha$ threshold \cite{fynbo05} contributes significantly  to the rate but the main path for the capture is through a high-lying $^{12}$C($2^+_2$) resonance,  which decays directly to the ground state of $^{12}$C \cite{diego11}.
Uncertainties in the properties of this $2^+_2$ state
have been recently addressed \cite{freer09}. It is important to note that this $2^+$ resonance has little influence at very low temperatures. Even if the triple-alpha {\em resonant} rate remains more uncertain than desired \cite{austin05},
it is far better understood that the {\em non-resonant} counterpart.

At temperatures below $T \approx 0.07$ GK, the 3 $\alpha$-particles, having no access to intermediate resonances, fuse directly.
The description of this process requires a good solution to the three-body scattering equation including Coulomb effects.
Recent work by the Kyushu group \cite{ogata09} makes use of the continuum discretized coupled channel (CDCC) method \cite{cdcc}
which expands the full three-body wavefunction in terms of the continuum states in the two-body subsystem (in this case $^8$Be).
In this method,  resonant and non-resonant processes are treated on the same footing.
The CDCC method has been widely used in nuclear reactions but the application to the triple-alpha capture is particularly challenging because
the charged particle reaction occurs at large distances and is dominated by the Coulomb interactions.
The CDCC results \cite{ogata09}, when compared to NACRE, show a 20 order of magnitude increase in the reaction rate at $T=0.02$ GK.
Soon after these results became available, numerous astrophysical studies were performed to understand the implications
of such a dramatic change. Stellar evolution calculations \cite{dotter09,suda11} demonstrate that the CDCC rate suppresses He flashes
in the asymptotic giant branch, and can result in the disappearance of the red giant phase in low and intermediate mass stars.
Studies of He ignition in accreting white dwarfs \cite{white-dwarfs} and accreting neutron stars
\cite{neutron-stars,peng-xrb} show that the CDCC rate is barely consistent with observations of Type Ia supernovae and type I X-ray bursts,
respectively.

The CDCC calculations in \cite{ogata09} represent the first attempt to take into account the non-resonant contribution
to the triple alpha cross section at very low relative energies, where it is expected to dominate.
The main drawback of the CDCC method is that, in general, the truncated
CDCC wavefunction does not have the correct asymptotic form for the scattering of three charged particles \cite{vasilevsky}.
It is known that the asymptotic form is naturally expressed in terms of hyperspherical coordinates \cite{vasilevsky}.
Recent attempts to solve the $\alpha+\alpha+\alpha$ problem by the Madrid-Aarhus collaboration \cite{alvarez07,diego10}
employ the hyperspherical adiabatic expansion method.  Limitations in the numerical techniques did not allow for
determining the rate below $T\approx 0.1$ GK. In this work we discuss another method based on the hyperspherical coordinates
which does provide improved accuracy and enables us to reach temperatures down to 0.01 GK.

An alternative, rather schematic, approach was recently proposed \cite{garrido11}. 
Inspired by the low-energy extrapolation of the sequential process, implemented in NACRE, Ref.~\cite{garrido11} proposed an extrapolation of the three-body Breit-Wigner form  for the  contribution to the three-body capture
that does not go through the $^8$Be resonance. This method is here referred as BW(3B).
Results for the triple-alpha reaction rate in \cite{garrido11} show a modification of the temperature dependence
of the rate at low temperature amounting to 7 orders of magnitude increase of the triple-alpha rate at 0.02 GK, when
compared to NACRE.

The works mentioned before \cite{ogata09,garrido11,diego10} approach the triple-alpha reaction as a three-body problem. Presently, a fully microscopic approach to the non-resonant triple-alpha capture is not feasible.
In this work we also follow the three-body approach, and expect microscopic corrections to this approach to be small compared to the three-body effects discussed in \cite{ogata09,garrido11}.

\medskip
%%%%%%%%%%%%%%%%%%%%%%%%%%%%%%%%%%%%%%%%%%%%%%%%%%%%%%%%%%%%%%%%%%%%%%%%%%%%%%%%%%%%%% method

The reaction rate $R_{abc}$ for $a+b+c \rightarrow D + \gamma$ at a given energy $E$
 can be related to the cross section for photo-dissociation $\sigma_{\gamma}$ through:
\begin{align}
R_{abc}(E)&= p! N_A^2 \; G_{abc,D} \; \frac{\hbar^3}{c^2} \frac{8\pi}{\left(\mu_{ab}\mu_{ab,c}\right)^{3/2}}
%\frac{2 (2J_D+1)}{(2J_a+1) (2J_b+1) (2J_c+1)}
 \; \frac{E^2_{\gamma}}{E^2} \; \sigma_{\gamma}(E_{\gamma})
\end{align}
with $E_{\gamma}$ the photon energy, $E$ the three-body energy, $\mu_{ab}$ and $\mu_{ab,c}$ the reduced
masses of  the $a+b$ and $(ab)+c$ systems, respectively, $p$ the number of identical particles,
$N_A$ Avogadro's number and $G_{abc,D}$ a statistical factor depending
on the spins of the nuclei $a,b,c$ and $D$ \cite{garrido11}.
The energies are related by $E_{\gamma}=E+|E_D|$, where $E_D$ is the energy of the bound final state of $D$
relative to the $3\alpha$ threshold. The relevant astrophysical quantity is the energy averaged reaction 
rate $\langle R_{abc}\rangle(T)$, obtained by integrating $R_{abc}(E)$ over the Maxwell-Boltzmann distribution.
%\begin{equation}
%\langle R_{abc}\rangle=\frac{1}{2}\frac{1}{(k_{B}T)^3}\int_{0}^{\infty}R_{abc}E^2e^{-\frac{E}{k_BT}}dE \;,
%\end{equation}
%where $k_B$ is the Boltzmann constant.

At low energies, the photo-dissociation process is usually dominated by a single transition of multipolarity $\lambda$, so
\begin{equation}
\sigma_{\gamma}=\frac{(2\pi)^3(\lambda {+}1)}{\lambda [(2\lambda{+}1)!!]^2}\left(\frac{E_{\gamma}}{\hbar c}\right)^{2\lambda -1}\frac{dB(E\lambda)}{dE}
\end{equation}
depending on the corresponding electromagnetic transition strength $dB(E\lambda)/dE$. For the triple-alpha reaction at low energies, the reaction
proceeds through a quadrupole transition from the $0^+$ continuum to the $2^+_1$ bound state in $^{12}$C.
%\begin{figure}[t!]
%\center
%\includegraphics[width=0.4\textwidth]{RateC12couplings-b.pdf}
%\caption{(Color online) Triple-alpha reaction rate: importance of long-range effects. We compare the full calculation (solid %line) with the results including only the diagonal Coulomb couplings (dot-dashed), and further adding off-diagonal Coulomb %couplings up to 90 fm (dashed).}
%\label{f1}
%\end{figure}

In order to calculate the transition strength $dB(E2)/dE$ at low relative energies, we need to solve the bound three-body problem
to obtain the  $2_1^+$ state and  to solve the full scattering three-body problem for the $0^+$ triple-alpha continuum.
To solve the three-body Faddeev equation, we use the hyperspherical harmonic (HH) method \cite{hh}. Hyperspherical coordinates (hyper-radius $\rho=\sqrt{x_i^2+y_i^2}$ and hyper-angle $\theta_i=\tan\frac{x_i}{y_i}$) are introduced
as functions of the scaled Jacobi coordinates $x_i,y_i$, with $i=1,2,3$ \cite{hh}. The HH method expands
the three-body wavefunction as sums of products of hyper-radial functions $\chi_{\gamma}(\rho)$ and angular functions $\Omega_{\gamma}(\theta,\hat{\vec x},\hat{\vec y})$ where $\gamma=\{i,K_i, l_{x_i}, l_{y_i}\}$.
This expansion introduces a new quantum number, the hyper-momentum $K_i$, and sufficiently many $K_i$ values need to be used to
obtain convergence. The orbital angular momenta $l_{x_i},l_{y_i}$ refer to the Jacobi coordinates $x_i,y_i$. Introducing the HH expansion in the Faddeev equations gives the set of coupled channels equations in the hyper-radius coordinate:
%\begin{equation}
%\!\left({-}\frac{\hbar^2}{2m}\!\left[\frac{d^2}{d\rho^2}{-}\frac{\Delta_i(\Delta_i{+}1)}{\rho^2}\right]{-}E\!\right)\chi_{\gamma}(\rho){=} \nonumber \\
%-\sum_{\gamma'}V_{\gamma \gamma'}(\rho)\chi_{\gamma'}(\rho),
%\label{hh-eq}
%\end{equation}
\begin{equation}
\left(\frac{\hbar^2}{2m}\left[\frac{d^2}{d\rho^2}{-}\frac{\Delta_i(\Delta_i{+}1)}{\rho^2}\right]{+}E\!\right)\chi_{\gamma}(\rho)=
\sum_{\gamma'}V_{\gamma \gamma'}(\rho)\chi_{\gamma'}(\rho),
\label{hh-eq}
\end{equation}
where $\Delta_{i}=K_i+3/2$ and $m$ is a scaling mass. 
The coupling potentials are defined as the sum of three pairwise interactions $V_{jk}(\rho,\theta)$ plus a three-body force $V_{3b}(\rho)$, integrated over all variables but $\rho$:
\begin{equation}
V_{\gamma \gamma'}(\rho)=
\langle \Omega_{\gamma}(\theta,\hat{\vec x},\hat{\vec y})|\sum_{k>j=1}^3\!\! V_{jk}+V_{3b}|\Omega_{\gamma'}(\theta,\hat{\vec x},\hat{\vec y})\rangle
\; .\label{coupl-eq}
\end{equation}
We take the same interactions for the three-body Hamiltonian as in \cite{alvarez07}. After minor adjustments, this reproduces the $^8$Be ground state, and the $^{12}$C $2^+_1$ bound state and $0^+_2$ Hoyle resonance. We explicitly take into account the symmetrization of the system, thereby reducing the set of coupled equations.

The solution for  the  $2_1^+$ bound state is obtained by solving Eq.\ (\ref{hh-eq}) for negative energy $E$. The numerical method is implemented in {\sc FaCE} \cite{face}, which is specifically designed for three-body bound states. The $0^+$ continuum states are found by using the R-matrix method in the HH basis (HHR), originally developed in the context of core+n+n scattering \cite{hh-rmatrix}, a problem with different symmetries and no Coulomb interactions. The HHR method generates an orthonormal basis set to solve Eq.\ (\ref{hh-eq}) in a box with hyper-radius $\rho_m$, by fixing the logarithmic derivative there. We expand $\chi_{\gamma}$
in terms of the HHR basis. Due to the long range Coulomb forces, numerical solutions are needed out to the very large distances ($\rho \gg 100$  fm) where it becomes safe to ignore off-diagonal couplings and matching to known Coulomb functions can be performed \cite{vasilevsky}. Extending the orthonormal basis to the required large $\rho$ would be impractical, so we use an R-matrix propagation technique \cite{light-walker} out to an asymptotic $\rho_{a} \gg \rho_m$. 
We divide the interval from $\rho_m$ to $\rho_{a} $ into sectors labelled $p$, 
the sizes of which are sufficiently small that the interactions within can be taken as constant.
The propagating functions $G^p$ are then given in terms of complex exponentials \cite{light-walker}, and used to propagate from sector $p{-}1$ to the next: 
\begin{align}
-R^p_{\gamma\gamma'}&=G^p_{\gamma\gamma'}(\rho_R^p,\rho_R^p)+\sum_{\alpha,\beta}G^p_{\gamma,\alpha}(\rho_R^p,\rho_L^p)\notag\\
&[R^{p-1}-G^p(\rho_L^p,\rho_L^p)]^{-1}_{\alpha,\beta}\; G^p_{\beta,\gamma'}(\rho_L^p,\rho_R^p)\; .\label{rexpand-eq}
\end{align}
This gives the R-matrix on the right side of the boundary $R^p$ from $R^{p-1}$
on the left. The coordinate subscripts $\rho_R$, $\rho_L$ imply evaluations on the right and the left side of the boundary respectively. While without Coulomb this propagation method was numerically stable, including the Coulomb interaction introduced numerical instabilities in the three-body scattering wavefunction, in the turning point region. To overcome these instabilities, we introduced screening in the off-diagonal potentials Eq.(\ref{coupl-eq}) using a Woods-Saxon multiplying factor $[1+\exp((\rho-\rho_{\mathrm{screen}})/a_{\mathrm{screen}})]^{-1}$. Our calculations use a sufficiently large screening radius such that $dB(E2)/dE$ for $E=0.05\textrm{--}0.5$ MeV becomes independent of $\rho_{\mathrm{screen}}$ and $a_{\mathrm{screen}}$.

\begin{figure}
\center
\vspace{-2cm}
\includegraphics[width=0.45\textwidth]{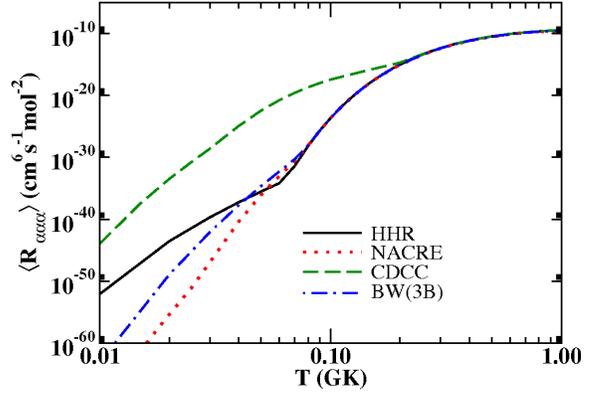}
\vspace{-3cm}
\caption{(Color online) Different evaluations of the triple-alpha reaction rate: comparing the Hyperspherical Harmonic R-matrix method (solid) with NACRE (dotted), CDCC (dashed) and the three-body Breit Wigner (dot-dashed).}
\label{f2}
\end{figure}

Convergence of $dB(E2)/dE$ was checked for all the relevant parameters. 
At relative energies  $\approx 0.1$ MeV, $dB(E2)/dE$ picks up contributions from the wave functions at hyper-radii $\rho=5\textrm{--}30$ fm. However, in order to get a reliable three-body scattering wavefunction, the matching to the asymptotic form needs to be performed at $\rho \gg \rho_{\mathrm{screen}}$. 
The results here presented include up to $K=26$ in the HH expansion, $50$ poles in the R-matrix expansion, an R-matrix box size of $50$ fm and R-matrix propagation out to $3000$ fm. In addition we use $\rho_{\mathrm{screen}}=800$ fm and $a_{\mathrm{screen}}=10$ fm
to screen the off-diagonal couplings. The only limiting factor for full convergence is the truncation in $K_{\max}$. We carefully studied the exponential behavior of $dB(E2)/dE$ as a function of $K_{\max}$, for the relative energy range $E=0.01\textrm{--}0.15$ MeV (the relevant region for rates below 0.07 GK). We conclude that our results with $K_{max}=26$ differ from those obtained from the extrapolation of $K_{\max} \rightarrow \infty$ by a factor of at most 2.  
Our energy averaged rates are plotted in Fig.\ref{f2} as a function of temperature (solid line). The factor of 2 error on the
rate at low energy cannot be seen given the scales in Fig.\ref{f2}.

%%%%%%%%%%%%%%%%%%%%%%%%%%%%%%%%%%%%%%%%%%%%%%%%%%%%%%%%%%%%%%%%%%%%%%%%%%%%%%%%%%%%%% results
\medskip

%To understand the importance of the long range couplings (Eq.\ \ref{rexpand-eq}), we show in Fig.\ \ref{f1} the reaction rate %obtained
%as a function of temperature for a few model scenarios. First we consider the full effect of both nuclear and Coulomb out to %$\rho=3500$ fm
%(solid). Next, the full result is compared to the rates obtained when neglecting off-diagonal potentials beyond $\rho=90$ fm %(dashed). The differences
%found below 0.07 GK demonstrate the need for the propagation method. Third, we show the results obtained when neglecting %off-diagonal couplings for all $\rho$, and retaining only the Coulomb diagonal parts (dot-dashed).
Our method to determine the three-body scattering state can be checked for the pure Coulomb case. 
In that case, no resonances are present 
in either the two-body or three-body systems and the capture is necessarily direct. For the pure
Coulomb case, the propagation technique is stable and we were able to unambiguously determine that 
no error is introduced by screening the off-diagonal couplings.
In addition, there is an analytic solution of Eq.\ (\ref{hh-eq}) for the pure diagonal Coulomb couplings 
which we use to test our implementation. The rate for diagonal Coulomb differs from the full rate by $10$ orders of magnitude, thus demonstrating the importance of including the Coulomb effects correctly. The off-diagonal Coulomb couplings are important, but their major contribution to the rates in the range $T=0.01\textrm{--}1.00$ GK occurs for $\rho<100$ fm.

In Fig.~\ref{f2} we compare our results (solid) with those obtained previously. The rate obtained from the full solution of the HHR equations agree with NACRE (dotted line) above 0.07 GK. Although there is a slight reduction of the rate below 0.07 GK, we find a pronounced increase of the rate for $T < 0.06$~GK, accompanied by a very different temperature dependence. In some ways, the results obtained assuming a rate extrapolated with a three-body Breit-Wigner form BW(3B) (dot-dashed line) \cite{garrido11} are qualitatively similar, but the BW(3B) treatment enhances the rate to a smaller degree. Although we do find a strong enhancement at low energy, it is not as strong as that seen in the CDCC result of \cite{ogata09} (long-dashed). While the deviation of the CDCC results from NACRE is seen all the way up to 0.11~GK, the HHR result agrees with NACRE for $T>0.07$~GK. This aspect turns out to be crucial in the repercussions for astrophysics.

The kink observed (Fig.~\ref{f2}) in the full rate around $T \approx 0.06\GK$ is a signature of two processes with different temperature dependence.  Above $T \approx 0.06\GK$ the process through the Holye state dominates, while below there is mostly non-sequential (direct) capture. The calculations in \cite{garrido11} exhibit this same feature. The temperature dependence is best illustrated by evaluating $d\ln \langle R_{\alpha\alpha\alpha} \rangle /d\ln T$, as shown in Table \ref{t1}. For $T < 0.06\GK$, the HHR rate has a much weaker temperature dependence than the NACRE rate, as is evident in Fig.~\ref{f2}. For $T > 0.06\GK$, the HHR rate follows that of NACRE, and their temperature sensitivities are basically equal (Table~\ref{t1}).

\begin{table}[b!]
\caption{Temperature sensitivity of triple-alpha rate}
\label{t1}
\begin{ruledtabular}
\begin{tabular}{ccc}
$T$ (GK) & \multicolumn{2}{c}{$d\ln \langle R_{\alpha\alpha\alpha}\rangle/d\ln T$} \\
 & \multicolumn{2}{c}{\hrulefill}\\
 & HHR & NACRE\\ \hline
0.01 & 34.1 & 56.5\\
0.02 & 23.3 & 45.5\\
0.04 & 18.5 & 47.7\\
0.08 & 51.7 & 48.3\\
0.16 & 24.4 & 24.4\\
\end{tabular}
\end{ruledtabular}
\end{table}

We have tested the sensitivity of the rate to the nuclear interaction and the three-body force by using the same Hamiltonian as \cite{ogata09} to calculate the $0^+$ continuum. The introduction of a three-body interaction was necessary to reproduce the relevant Hoyle state. This modified Hamiltonian increases the rates at  $T=0.02\,\mathrm{GK}$ by 4 order of magnitude, compared to those obtained with the Ali-Bodmer interaction \cite{ali-bodmer}. However, while the Ali-Bodmer  interaction (used in \cite{alvarez07} and also here),  
reproduces the $\alpha\textrm{-}\alpha$ phase shifts, that of \cite{ogata09} does not.

%%%%%%%%%%%%%%%%%%%%%%%%%%%%%%%%%%%%%%%%%%%%%%%%%%%%%%%%%%%%%%%%%%%%%%%%%%%%%%%%%%%%%% astro
\medskip

\begin{figure}
\center
\includegraphics[width=0.45\textwidth]{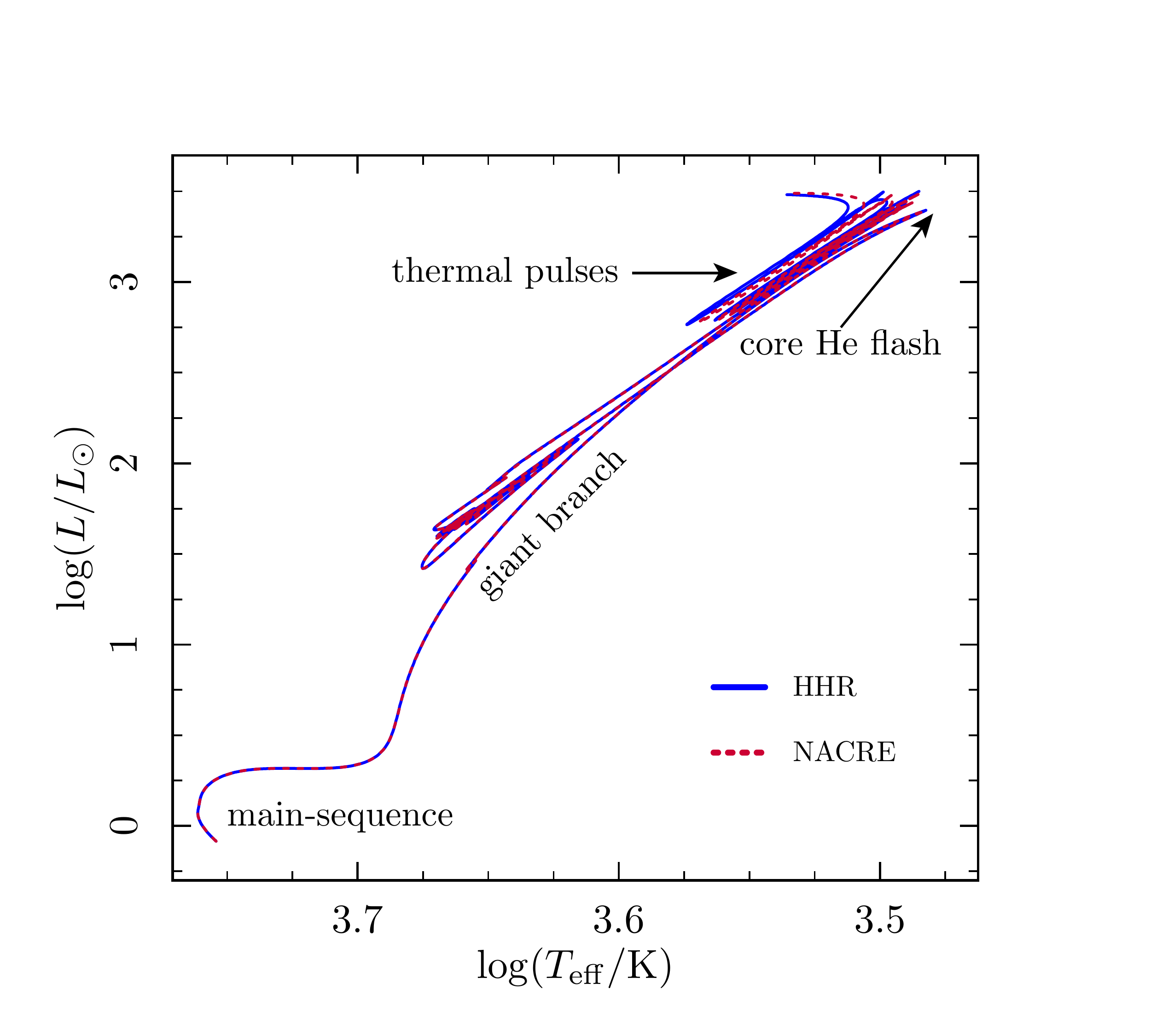}
\caption{(Color online) Evolutionary track (luminosity vs.\ surface effective temperature) of a
one solar mass star with solar composition, for the HHR rate (\emph{solid line}) and the NACRE
rate (\emph{dashed line}).  The evolution is identical for both from when  H fuses to He in the
core (``main sequence'') thorough the formation of a degenerate He core (``giant branch'') and
the ignition of He in the core (``core He flash'').  Small differences are seen when thermally
unstable H and He burning occurs in a shell about a degenerate C/O core ("thermal pulses"), but
there is no difference in the final white dwarf's mass or composition.}
\label{f3}
\end{figure}

Next we study the effect of the strong enhancement of the triple-alpha rate at low temperature on stellar evolution. We use the \textsc{mesa} (Modules for Experiments in Stellar Astrophysics)  code library \cite{mesa}, but replace the NACRE rate by the HHR rate for $T < 0.1 \GK$.  We explore the evolution of progenitors with mass $M=0.8\Msun$, $1.0\Msun$, and $1.25\Msun$, all with a standard solar composition, using the setups described in Ref.~\cite[\S~7.1]{mesa}.
In general, He burning occurs in low-mass stars only for $T > 0.07\GK$, and therefore the changes to the stars are minimal. Figure~\ref{f3} illustrates the post main sequence evolutionary track, in the luminosity--effective temperature plane, for a star of mass $1 \Msun$.  Tracks for the standard NACRE rate (dashed line) and the HHR rate (solid line) are shown.  Both stars go through a red-giant phase until He ignites in the semi-degenerate core, unlike the case calculated with the CDCC rate, for which the red-giant phase vanishes \cite{dotter09}. 
Only after cessation of core He burning and the onset of H and He thermal shell flashes do small differences appear (shown in Fig.\ref{f3}): the enhanced HHR rate at $T < 0.06\GK$ tends to produce a slightly larger convective zone during helium shell flashes. Both rates lead to final white dwarfs with identical masses and carbon-oxygen ratios.  The evolutionary tracks for a $1.25\Msun$ star behave similarly to the $1\Msun$ case, while the tracks for the $0.8\Msun$ star are identical for both the NACRE and HHR rates.
Further
astrophysical studies to understand the implications of the new rate in explosive scenarios will be carried out in the near future.

%%%%%%%%%%%%%%%%%%%%%%%%%%%%%%%%%%%%%%%%%%%%%%%%%%%%%%%%%%%%%%%%%%%%%%%%%%%%%%%%%%%%%% conclusions

In summary, we have developed a framework combining the Hyperspherical method, the R-matrix
expansion, and the R-matrix propagation and screening technique (named here HHR) to calculate the 
triple-alpha reaction rate at low temperatures, $T < 0.1\GK$. Our results show a strong 
enhancement of the rate below $0.06\GK$, while reproducing the NACRE rate at higher temperatures.
A longer publication with a detailed account of the method and further analysis of the reaction dynamics
is underway.

\begin{acknowledgments}
We thank Richard Cyburt, Ron Johnson, Akram Mukhamedzhanov and Chuck Horowitz for useful discussions during
this project.
This work was supported by the National Science Foundation grant PHY-0800026 and the Department of Energy under contracts DE-FG52-08NA28552
and DE-SC0004087.
This work was performed under the auspices of the U.S. Department of
Energy by Lawrence Livermore National Laboratory under Contract
DE-AC52-07NA27344.
\end{acknowledgments}


\begin{thebibliography}{12}
\bibitem{hoyle54} F. Hoyle, Astrophys. J. Suppl. Ser. 1 (1954) 121.

\bibitem{dunbar53} D.N.F. Dunbar, R.E. Pixley, W.A. Wenzel and W. Whaling, Phys. Rev. 92, 649 (1953).

%\bibitem{caugh-fow1} W.A. Fowler, G.E. Caughlin, B.A. Zimmerman, Annu. Rev. Astron. Astrophys. 5, 525 (1967).

\bibitem{nacre} C. Angulo {\it et al.}, Nucl. Phys. A 656, 3 (1999).

\bibitem{ogata09} K. Ogata, M. Kan and M. Kamimura, Prog. Theor. Phys. 122, 1055 (2009).

\bibitem{garrido11} E. Garrido, R. de Diego, D.V. Fedorov and A.S. Jensen, Eur. Phys. J A 47, 102 (2011).

\bibitem{fynbo05} H.O.U. Fynbo {\it et al.}, Nature 433, 136 (2005).

\bibitem{diego11} R. de Diego, E. Garrido, D.V. Fedorov and A.S. Jensen, Phys. Lett. B 695, 324 (2011).

\bibitem{freer09} M. Freer {\it et al.}, Phys. Rev. C 80, 041303 (2009).

\bibitem{austin05} Sam M Austin, Nucl. Phys. A 758, 375c (2005).


\bibitem{cdcc} N. Austern, Y. Iseri, M. Kamimura, G. Rawitscher and M. Yahiro, Phys. Rep. 154 (1987) 125; M. Yahiro, N. Nakano, Y. Iseri and M. Kamimura, Prog. Theo. Phys. 67 (1982) 1464; Prog. Theo. Phys. Suppl. 89, 32 (1986).


\bibitem{dotter09} A. Dotter and B. Paxton, Astron. Astrophys. 507, 1617 (2009).

\bibitem{suda11} T. Suda, R. Hirschi and M. Fujimoto, Astrophys. J. 741, 61 (2011).

\bibitem{white-dwarfs} M. Saruwatari and M. Hashinoto, Prog. Theor. Phys. 124, 925 (2010).

\bibitem{neutron-stars} Y. Matsuo {\it et al.}, arxiv:1105.5484v2

\bibitem{peng-xrb} F. Peng, and C.~D. Ott, Astrophys.~J.\ 725, 309 (2010).

\bibitem{vasilevsky} V. Vasilevsky, A.V. Nesterov, F. Arickx and J. Broeckhove, Phys. Rev. C 63, 034606 (2001).


\bibitem{alvarez07} R. Alvarez-Rodriguez, E. Garrido, A.S. Jensen, D.V. Fedorov, H.O.U. Fynbo, Eur. Phys. J. A 31, 303 (2007).

\bibitem{diego10} R. de Diego, E. Garrido, D.V. Fedorov and A.S. Jensen, Eur. Phys. Lett. 90, 52001 (2010).



\bibitem{maris} Pieter Maris, UNDEF collaboration meeting, East Lansing June 2011.

\bibitem{pieper} S. Pieper, UNDEF collaboration meeting, East Lansing June 2011.

\bibitem{neff07} M. Chernykh, H. Feldmeier, T. Neff, P. von Neumann-Cosel and A. Ritcher, Phys. Rev. Lett. 98, 032501 (2007).




\bibitem{hh} L.M. Delves, Nucl. Phys. 9, 391 (1959); Nucl. Phys. 20, 275 (1960).

\bibitem{face} I.J. Thompson, F.M. Nunes, B.V. Danilin, Computer Physics Communication 161, 87 (2004).

\bibitem{hh-rmatrix} 
I.J. Thompson, B.V. Danilin, V.D. Efros, J.S. Vaagen, J.M. Bang and M.V. Zhukov, Phys. Rev. C 61, 24318 (2000).


\bibitem{light-walker} J.C. Light, R.B. Walker, J. Chem. Phys. 65, 4272 (1976).

\bibitem{ali-bodmer} S. Ali and A.R. Bodmer, Nucl. Phys. 80, 99 (1966).

%\bibitem{fedorov96} D.V. Fedorov, A.S. Jensen, Phys. Lett. B 389, 631 (1996).

\bibitem{mesa} B. Paxton, L. Bildsten, A. Dotter, F. Herwig, P. Lesaffre and F. Timmes, Astrophys. J. Suppl.\ Ser.\ 192, 3 (2011) \url{http://mesa.sourceforge.net/} (version 3635).

\end{thebibliography}
\end{document}